\documentclass[preprint, 12pt, proof, authoryear]{elsarticle}
\usepackage[left=1in, right=1in, top = 1.5in, bottom = 1.5in]{geometry}
\usepackage[utf8]{inputenc}
\usepackage{csquotes}
\usepackage[all]{xy}
\usepackage{amsmath}
\usepackage{csquotes}
\usepackage{subfig}
\usepackage{enumerate}
\usepackage{graphicx}
\usepackage{enumerate}
\usepackage{color}
\usepackage{mwe}
\usepackage[titletoc]{appendix}
\usepackage{centernot}
\usepackage{url,hyperref,xspace}
\usepackage[ruled,vlined]{algorithm2e}
\usepackage{soul}
\usepackage{xcolor}
\usepackage{mathtools}
\usepackage{footnote}
\usepackage{multirow}
\usepackage{tablefootnote}
\usepackage{caption}
\usepackage{adjustbox}
\usepackage{booktabs}
\usepackage[normalem]{ulem}
\usepackage{cancel}
\usepackage{titlesec}
\usepackage{titletoc}
\usepackage{natbib}
\setcitestyle{ sort&compress}
\usepackage{amssymb,centernot, amsfonts}
\usepackage{amsthm}

\makesavenoteenv{tabular}
\usepackage[flushleft]{threeparttable}
\def\@centernot#1#2{%
  \mathrel{%
    \rlap{%
      \settowidth\dimen@{$\m@th#1{#2}$}%
      \kern.5\dimen@
      \settowidth\dimen@{$\m@th#1=$}%
      \kern-.5\dimen@
      $\m@th#1\not$%
    }%
    {#2}%
  }%
}
\makeatother

\newcommand{\independent}{\perp\mkern-9.5mu\perp}

\usepackage{tikz}
\usetikzlibrary{arrows.meta,shapes}
\usetikzlibrary{arrows,shapes.arrows,shapes.geometric,shapes.multipart,
decorations.pathmorphing,positioning,shapes.swigs,}

\usepackage{setspace}
\usepackage[resetlabels]{multibib}
\newcites{SM}{Appendix References}
\newcites{Main}{References}

\newtheorem{proposition}{Proposition}
\newtheorem*{proposition*}{Proposition}

\MakeOuterQuote{"}\EnableQuotes
\DeclareUnicodeCharacter{00A0}{ }

\usepackage[final]{changes}

\makeatletter
\def\ps@pprintTitle{%
 \let\@oddhead\@empty
 \let\@evenhead\@empty
 \def\@oddfoot{}%
 \let\@evenfoot\@oddfoot}
\makeatother

\begin{document}

\begin{frontmatter}
\title{Rejoinder to  ``Perspectives on \replaced{`harm'}{harm} in personalized medicine - an alternative perspective''}

\author{Aaron L. Sarvet  \corref{cor1}}
\author{Mats J. Stensrud}

\address{Department of Mathematics, École Polytechnique Fédérale de Lausanne, Switzerland}

\begin{abstract}
In our original article \citep{sarvet2023perspectives}, we examine twin definitions of ``harm'' in personalized medicine: one based on predictions of individuals' unmeasurable response types (\textit{counterfactual} harm), and another based solely on the observations of experiments (\textit{interventionist} harm). In their commentary, \citet{mueller2023perspective} (MP) read our review as an argument that ``counterfactual logic should $[\dots]$ be purged from consideration of harm and benefit'' and ``strongly object $[\dots]$ that a rational decision maker may well apply the interventional perspective to the exclusion of counterfactual considerations.'' Here we show that this objection is misguided. We analyze MP's examples and derive a general result, showing that determinations of harm through \textit{interventionist} and \textit{counterfactual} analyses will always concur. Therefore, individuals who embrace counterfactual formulations and those who object to their use will make equivalent decisions in uncontroversial settings. 
\end{abstract}

\end{frontmatter}

\newpage

\section*{Introduction} \label{sec:intro}
\citetMain{mueller2023perspective} (herein, MP) claim that a ``\textit{counterfactual} approach is vital for effective decision-making policies.'' They particularly caution against an \textit{interventionist} approach that bases decisions solely on parameters of hypothetical experiments. MP support their claim with an example, where ostensibly only a \textit{counterfactual} analysis would detect a harmful treatment.  

MP's argument is flawed because \textit{interventionist} and \textit{counterfactual} approaches are compared on uneven grounds: relevant and available information in non-experimental data is exclusively applied in their \textit{counterfactual} analysis. In reconsidering their example, we show that MP's concerns are resolved when this information is also used in an \textit{interventionist} approach. 

The strategy that resolves MP's concern extends beyond their example; we give new theoretical results on the concordance of \textit{counterfactual} and \textit{interventionist} analyses. Using MP's minimal assumptions, we show that \textit{counterfactual} and \textit{interventionist} analyses will always concur in their determination of harm. Thus, there cannot exist any example where \textit{counterfactual} logic is strictly necessary for this aim.

\section*{MP's example} \label{sec: example}

\subsection*{Preliminaries}
MP consider a hypothetical treatment for a deadly disease. Patients were randomly treated ($A=1$) or untreated ($A=0$) in an experiment, and patient deaths ($Y$) were recorded after one year. MP suppose that an analyst has access to this experimental data, and also non-experimental data wherein individuals took treatment as they would naturally. 
Formally, we let $P_0$ and $P_1$ denote the observable parameters of the experimental and non-experimental data, respectively. We consider the raw data presented in Table 3 of \citetMain{mueller2023personalized}. To reduce clutter, we focus on results among men, as it is in this group where MP claim a distinction between \textit{interventionist} and \textit{counterfactual} analyses.

We consider definitions of harm detection consistent with MP's analysis and discussion. Harm is detected in an \textit{interventionist} analysis when the observable parameters imply \textit{interventionist} harm, i.e.,  that more people are expected to die when treated than untreated for some group with relevant feature $X=x$, $$\Pr(Y^{a=1}=1 \mid X=x) - \Pr(Y^{a=0}=1\mid X=x) > 0.$$ 
Alternatively, harm is detected in a \textit{counterfactual} analysis when the observable parameters imply a positive probability of \textit{counterfactual} harm, i.e., that there is a positive probability of the \textit{principal stratum} of individuals who would be ``killed'' by treatment, $$\Pr(Y^{a=1}=1, Y^{a=0}=0) > 0.$$ 

The left hand side of the inequalities do not need to be point-identified in order to make a determination in either analysis: a lower bound greater than 0 would imply some amount of harm in either case.  A lower bound is \textit{sharp} with respect to a set of observable parameters if it is the \textit{highest-possible} lower bound for those parameters. If a sharp lower bound is less than or equal to 0, then those parameters cannot possibly imply that harm occurred. When the value of a causal estimand is point identified, then trivially this is also the sharp lower bound. 

\subsection*{The natural treatment value}

MP's results are fundamentally misleading because they fail to leverage a relevant feature measured in the data. Due to this oversight, MP conclude that the non-experimental data are useless in the interventional analysis. The relevant feature is the patient's natural treatment value, i.e., their intended treatment, $A^*$, and is implicitly used in MP's \textit{counterfactual} analysis.  In a non-experimental setting $A^*$ is measured; no intervention is made and the treatment a patient actually receives, $A$, is indeed equal to this \textit{natural value}. In experimental settings, however, the patients' natural treatment intentions may be subverted by the experimental design, and thus are rarely measured outside of special settings \citepMain{long2008causal, knox2019design}. The natural treatment value has an extended history in causal inference \citepMain{robins2004effects, robins2006comment, haneuse2013estimation, richardson2013single, young2014identification} and is receiving increasing attention in epidemiology \citepMain{hoffman2024introducing, sarvet2020graphical}. 

When experimental and non-experimental data are available,  \textit{interventionist} harm is indicated by the average treatment effects (ATE) among those who did and did not intend to take treatment naturally, $$\Pr(Y^{a=1}=1 \mid A^*=a') - \Pr(Y^{a=0}=1\mid A^*=a'),$$ for $a' \in \{0,1\}$. These ATEs have been historically referred to as the average treatment effect in the treated ($A^*=1$, the ATT) and the untreated ($A^*=0$, the ATU) \citepMain{bloom1984accounting, heckman1990varieties}.

\subsection*{Results of a \textit{counterfactual} analysis}
Table \ref{tab:pearl} gives sharp lower bounds on causal parameters indicating harm (the rows) in two sets: the bounds implied by observable parameters of the experimental data alone, $P_0$; and the bounds implied when additionally using the non-experimental data, $P_1$. The first row gives sharp lower bounds on the probability of \textit{counterfactual} harm, using results in \citetMain{tian/pearl:probcaus}. The experimental data alone imply a sharp lower bound of $0$, and so MP cannot say that any \textit{counterfactual} harm occurred with only these data. With additional non-experimental data, a sharp lower bound of 0.21 is obtained, and so they conclude that \textit{counterfactual} harm occurred. Indeed, in MP's example the probability of \textit{counterfactual} harm is point-identified by this value, even though point identification is not in general guaranteed when non-experimental data are additionally used. We show that point identification arises in this example due to hidden determinisms that are revealed in a full \textit{interventionist} analysis. 

\subsection*{Results of an \textit{interventionist} analysis}

The second row of Table \ref{tab:pearl} gives sharp lower bounds on the marginal ATE. The experimental data alone imply a sharp lower bound of $-0.28$. As this ATE is indeed identified by the experimental data, then this bound cannot logically be improved by any additional data. Thus MP end their \textit{interventionist} analysis and, contrary to their \textit{counterfactual} analysis, cannot conclude that any \textit{interventionist} harm occurred.

However, MP neglect to examine the conditional ATEs, among patients who did intend ($A^*=1$) and did not intend ($A^*=0$) to take treatment. 
The sharp lower bounds for these effects are provided in the bottom two rows of Table \ref{tab:pearl}. As patients' treatment intentions are not measured in the experimental data, they alone say nothing about these conditional ATEs: the sharp lower bounds for these parameters are thus -1. In contrast, the addition of non-experimental data point-identifies these effects. Following \citetMain{stensrud2022optimal}, who build on classical results for the identification of the ATT \citepMain{robins2007causal,dawid2022can,geneletti2011defining,bareinboim2015bandits}, we detect \textit{interventionist} harm among those who did not intend to take treatment in the non-experimental data ($A^*=0$). Consideration of these conditional ATEs not only resolves disagreement between the \textit{interventionist} and \textit{counterfactual} analyses; it also indicates precisely, via a measurable variable, in which group \textit{interventionist} harm occurs ($A^*=0$). It is also precisely in this group that \textit{counterfactual} harm occurs, $\Pr(Y^{a=1}=1, Y^{a=0}=0 \mid A^*=0)>0$;  alternatively, among those who intended to be treated, the probability of \textit{counterfactual} harm is exactly 0, $\Pr(Y^{a=1}=1, Y^{a=0}=0 \mid A^*=1)=0$. This argument shows that MP missed an opportunity to improve their \textit{counterfactual} analysis and subsequent decisions or policies.

These results are not a convenient coincidence that occurred in MP's example. In the next section, we show that they follow from general relations between \textit{counterfactual} and \textit{interventionist} analyses.

\begin{table}
    \centering
    \begin{tabular}{|l | l | c | c | c| }
    \hline
  \multirow{2}{*}{\textbf{Approach}} & \multirow{2}{*} {\textbf{Estimand}} & \multicolumn{2}{c|}{\textbf{Sharp lower bound}} & \multirow{2}{*} {\textbf{Harm?}}  \\ \cline{3-4}
  &  & $P_0$ only & $P_0$ \& $P_1$ & \\
    \hline
   \textit{Counterfactual}    & $\Pr(Y^{a=1}=1, Y^{a=0}=0)$                            & 0     & $\mathbf{0.21}$ & Yes\\ \hline
  \multirow{3}{*}{\textit{Interventionist}}   & $\mathbb{E}(Y^{a=1}) - \mathbb{E}(Y^{a=0})$                      & -0.28 & -0.28 & \multirow{3}{*}{{Yes}} \\
     & $\mathbb{E}(Y^{a=1}\mid A^*=1) - \mathbb{E}(Y^{a=0}\mid A^*=1)$ & -1    & -0.70 & \\
     & $\mathbb{E}(Y^{a=1}\mid A^*=0) - \mathbb{E}(Y^{a=0}\mid A^*=0)$ & -1    & $\mathbf{0.70 }$ &   \\ \hline
    \end{tabular}
    \caption{Sharp lower bounds implied by the experimental data alone $(P_0)$, or with the addition of non-experimental data $(P_1)$ in MP's example. Positive values (in \textbf{bold}) indicate harm.}
    \label{tab:pearl}
\end{table}

\section*{Formal results relating \textit{counterfactual} and \textit{interventionist} analyses}

Let $\Phi_L$ denote the sharp lower bound on the probability of \textit{counterfactual} harm implied by the observable experimental parameters $P_0$. Let $\Phi^*_L$ give that sharp bound implied by the addition of non-experimental data $P_1$. Similarly, let $\Psi_L$ denote the sharp lower bound on the marginal ATE implied by $P_0$, and $\Psi_L^*({a'})$ denote the sharp lower bound on the conditional ATE given $A^*=a'$ implied by the addition of $P_1$. Then we have the following proposition.
    
\begin{proposition}  \label{prop: 1} 
Suppose experimental and non-experimental data can be properly fused. Then,
\begin{itemize}
    \item [(1)] $\Phi_L > 0$ if and only if $\Psi_L > 0$, and
    \item [(2)] $\Phi_L^* > 0$  if and only if $\Psi^*_L({a'}) >0$ for some $a'\in \{0,1\}$.
\end{itemize}
\end{proposition}

The implication of this Proposition is important: the cases when a \textit{counterfactual} analysis would detect harm exactly coincide with those cases when an \textit{interventionist} one would do so. In this sense, \textit{counterfactual} logic is entirely unnecessary.

The proofs of all our formal results, including Proposition \ref{prop: 1}, are given in the web appendix.

\subsection*{A remark on point identifying \textit{counterfactual} harm}

The probability of \textit{counterfactual} harm is point identified only in special cases. It is well known that the probability of \textit{counterfactual} harm is point identified by experimental data alone if and only if the probability of death is 1 or 0 in one of the experimental arms, what \citetMain{dawidharm} call ``a most unusual state of affairs''. This was not the case in MP's example; they only achieve point identification with the addition of the non-experimental data.
However, we do observe a different unusual state of affairs: those who \textit{did not intend to be treated} ($A^*=0$) would certainly die under treatment, whereas those who \textit{did intend to be treated} ($A^*=1$)  would certainly die without it; apparently, patients in the non-experimental data of MP's example were exceptionally adept at avoiding treatment conditions that result in certainly death.

 The following proposition from \citetMain{dawidharm} shows that such unusual determinisms will generally be the rule whenever probabilities of \textit{counterfactual} harm are identified as in MP's example. 

\begin{proposition}[Remarks 1 \& 2 in \citetMain{dawidharm}]     \label{prop: 2}
Suppose the conditions of Proposition \ref{prop: 1} hold, and let $\Phi_U$ and $\Phi_U^*$ denote the sharp upper bounds on the probability of \textit{counterfactual} harm, analogous to  $\Phi_L$ and $\Phi_L^*$. Then:
\begin{itemize}
    \item [(i)] $\Phi_L=\Phi_U$ if and only if $\Pr(Y^a=1) \in \{0,1\}$ for some $a\in \{0,1\}$;
    \item [(2)] $\Phi^*_L=\Phi^*_U$ if and only if both
    \begin{itemize}
        \item [(i)] $\Pr(Y^a=1 \mid A^*=1) \in \{0,1\}$ for some $a,\in \{0,1\}$, and
        \item [(ii)] $\Pr(Y^{a^\circ}=1 \mid A^*=0) \in \{0,1\}$ for some $a^{\circ}\in \{0,1\}$.
    \end{itemize} 
\end{itemize}
\end{proposition}

Utilizing Propositions \ref{prop: 1} and \ref{prop: 2}, we can make more informative statements about the probability of \textit{counterfactual} harm than MP made in their own example: among those who intended to be treated, the probability of \textit{counterfactual} harm is identified to be precisely 0, indicating to a \textit{counterfactual} analyst that the treatment is surely safe within a known group defined by a measurable covariate. This is not a coincidence; whenever the addition of non-experimental data would point identify a positive probability of \textit{counterfactual} harm when the experimental data alone would not, then either those who did or did not intend to take treatment will have probability 0 of \textit{counterfactual} harm. Proposition \ref{prop: 3} in the Web Appendix formalizes and generalizes this result.
 
As we have argued, harm determination does not require point identification of the probability of \textit{counterfactual} harm; sharp lower bounds are sufficient. In contrast, bounds pose unique challenges for personalized medicine, as we discussed in \citetMain{sarvet2023perspectives}[Section 3].

\section*{Conclusion \& metaphysical clarification}

MP argue that \textit{counterfactual} logic is necessary and liken its role to the use of ``imaginary numbers'' in mathematics and engineering. We find it ironic that MP would make an analogy with such a term, being a pejorative coined by René Descartes in reference to the ``unreality'' of objects defined by the square-root of negative 1.

Imaginary numbers are surely useful tools in many areas of mathematics, from pure to applied, and arguably played remarkable roles in the discovery of important results. However, MP go further and claim that imaginary numbers are ``indispensable even in the analysis of real quantities.'' But the use of imaginary numbers in algebra and applied sciences has historically been met with trepidation; wherever imaginary numbers have appeared essential, there have been complementary efforts to provide alternative proofs that did not deal in imaginary terms, from the French mathematician Francois Vi\`ete's 16th century trigonometric solution to the roots of cubic functions \citepMain{viete1970opera}, to contemporary reformulations of the  famous ``Schr\"odinger equation'' in quantum mechanics \citepMain{callender2023quantum}. More generally, the proper and improper use of ``imaginaries'' in formal mathematical theories has been a major topic of investigation at the intersection of mathematics and analytic philosophy, prominently explored in the works of David Hilbert and Edmund Husserl \citepMain{majer1997husserl}. 

The preeminent 19th-century British astronomer George Airy once remarked ``I have not the smallest confidence in any result which is essentially obtained by the use of imaginary symbols'' \citepMain{airy1864supplement}.  In causal inference, this sensibility has been championed by contemporary statistician Philip Dawid, who has for the past decades, demonstrated with colleagues the non-necessity of potential outcomes for many tasks in causal inference, see \citepMain{dawid2021decision} for a review. 
Echoing Descartes, Dawid once pejoratively likened principal strata to an ``imaginary can-opener'' \citepMain{dawid2012imagine}, and asked:

\begin{displayquote}
    ``But what if there is no real can-opener -- no real-world pre-treatment variable corresponding to the fictitious principal stratum? There is then no way of determining which principal stratum an individual belongs to. How can a principal stratum analysis then tell us anything relevant about the real world?''
\end{displayquote}

In showing the general concordance of \textit{interventionist} and \textit{counterfactualist} analyses in determination of harm, we support the projects of Dawid and others, who would remain skeptical of results whose meaning depended on fundamentally unobservable quantities. Such skepticism should not prevent anyone from using \textit{counterfactual} logic or language. In fact, we use \textit{counterfactual} language extensively in our own work, along the lines of pioneers in causal inference who advocate for ``single world'' estimands and assumptions \citepMain{robins2010alternative,richardson2013single}. However, we do believe that careful attention is needed when \textit{counterfactual} approaches based on principal strata yield discrepant answers, especially when human lives are at stake.

\bibliographystyleMain{plainnat}
\bibliographyMain{refs}

\clearpage
\renewcommand{\appendixname}{Web Appendix}

\begin{appendices}
\doublespacing
\setcounter{page}{1}
\renewcommand{\thesection}{\arabic{section}}
\titlelabel{Web Appendix \thesection.\quad}



\section{A formal result on the point identification of counterfactual harm}

Define $\Phi^*_{L}(a')$ and  $\Phi^*_{U}(a')$ as the sharp lower and upper bounds on the probability of \textit{counterfactual} harm among those with $A^*=a'$, i.e., $\Pr(Y^{a=1}=1, Y^{a=0}=0 \mid A^*=a')$. Define $\beta^*_{L}(a')$ and  $\beta^*_{U}(a')$ as the sharp lower and upper bounds on the probability of \textit{counterfactual} benefit among those with $A^*=a'$, i.e., $\Pr(Y^{a=1}=0, Y^{a=0}=1 \mid A^*=a')$. Then we have the following result:

\begin{proposition}    \label{prop: 3}
Suppose the conditions of Proposition \ref{prop: 1} hold, and suppose $\Phi^*_L=\Phi^*_U>0$ or $\Phi_L=\Phi_U>0$. Then, one of the following equalities holds for each $a'$,

\begin{itemize}
    \item [(1)] $\beta^*_{L}(a') = \beta^*_{U}(a') = 0$, or
    \item [(2)] $\Phi^*_{L}(a') = \Phi^*_{U}(a') = 0$.
\end{itemize}
\end{proposition}

To fix ideas about the proposition, suppose that an analyst point identifies a (marginal) positive probability of \textit{counterfactual} harm, with either experimental data alone or additionally with non-experimental data. Then, conditional on an individual's treatment intention, we point identify that either no \textit{counterfactual} harm could possibly occur or that no \textit{counterfactual}  benefit could possibly occur. Suppose further that a \textit{counterfactual} investigator will certainly treat a patient when no \textit{counterfactual} harm could possibly occur, and that they will certainly \textit{not} treat a patient when no \textit{counterfactual} benefit could possibly occur. Then, in settings like MP's example where the probability of \textit{counterfactual} harm is point-identified marginally, this investigator would always make treatment decisions with confidence, so long as they consider a patient's natural treatment intention as a legitimate covariate in their dynamic regime.

\section{Proofs}

We introduce an additional variable $R$, a binary indicator of participation within a controlled experiment on $A$. As in Web Appendix 2 of \citetSM{sarvet2023perspectives}, we define an elaborated model $\mathcal{M}$, which clarifies our definitions of \textit{experimental} data, \textit{non-experimental} data, and what we mean when we say that the data sources \textit{can be properly fused}.

\begin{itemize}
    \item[A1.] (\textbf{Distributional consistency.})
 $\mathbb{E}[Y^{a} \mid A=a, R=r] = \mathbb{E}[Y \mid A=a, R=r]$ for all $r,  a$.
    \item[A2.] (\textbf{Experimental data.})  $Y^a \independent A \mid  R=1$ for all $a$.
    \item[A3.] (\textbf{Non-experimental data.}) $\Pr(A=A^* \mid R=0)=1$.
    \item[A4.] (\textbf{Fusing data.}) $(Y^{a=1}, Y^{a=0}, A^*)  \independent R $.
\end{itemize}
We discuss these assumptions in Web Appendix 2 of \citetSM{sarvet2023perspectives}. The arguments in all of the proofs are analogous when terms additionally condition on a baseline covariate $L$, like ``sex'' in MP's example. However, we have omitted $L$ from all our arguments to avoid clutter.

First, we re-state the following result from \citetSM{sarvet2023perspectives}, where we let $\Psi^*_U(a^{\dagger})$ denote the sharp upper bound on the CATE for $A^*=a^{\dagger}$, implied by $P_0$ and $P_1$:
\begin{proposition*}[Proposition 4 from the Web Appendix 2 of \citetSM{sarvet2023perspectives}]
    Suppose the conditions of Proposition \ref{prop: 1} hold. Then $\Phi_L^*>\Phi_L$ if and only if $\Psi_U^*(a^{\dagger})<0<\Psi_L^*(a^{\ddagger})$ for some $a^\dagger \neq a^\ddagger$.
\end{proposition*}
This proposition states that the sharp lower bound on the probability of \textit{counterfactual harm} will improve upon the addition of non-experimental data if and only if the signs of the CATEs conditional on $A^*$ are strictly opposite.

\subsection*{Proof of Proposition 1}
Property (1) of Proposition 1 follows from the classical results that the ATE is equal to $\Pr(Y^{a=1}=1, Y^{a=0}=0)  - \Pr(Y^{a=1}=0, Y^{a=0}=1) $ when $Y$ and $A$ are binary variables. 

Property (2) follows from Propositions 1 and 4 of the Web Appendix in \citetSM{sarvet2023perspectives}. In particular, Proposition 4 of \citetSM{sarvet2023perspectives} states directly that, if the lower bound on the probability of \textit{counterfactual} harm improves with the addition of non-experimental data, then one and only one ATE conditional on $A^*$ will be positive. Proposition 1 of \citetSM{sarvet2023perspectives} implies that conditional ATEs are identified, and so the sharp lower bounds are equal to the true conditional ATEs. We also use the fact that the marginal ATE is a convex combination of the conditional ATEs, and thus the sign of at least one of the conditional ATEs will agree with the sign of the marginal ATE. 

Suppose that the lower bounds do not improve with the addition of non-experimental data. If the lower bound was 0, then the marginal ATE was negative or zero and thus both (sharp lower bounds on the) conditional ATEs are also negative or zero. If the lower bound was positive, then the marginal ATE was positive, and similarly both (sharp lower bounds on the) conditional ATEs are also positive. Therefore, the ``if and only if'' statement holds when the lower bounds do not improve. Suppose that the lower bounds do improve, then the improved sharp lower bound can only be positive. According to Proposition 4 of \citetSM{sarvet2023perspectives}, however, we know that one conditional ATE will be positive. This concludes the proof.

\subsection*{Proof of Proposition 3}
It must be the case that $\Phi_L^*=\Phi_L$ or $\Phi_L^*>\Phi_L$, since a sharp lower bound cannot decrease with additional data.

If $\Phi_L^*=\Phi_L$, then we know that $\Pr(Y^{a=1}=1 \mid A^*=a') - \Pr(Y^{a=0}=1 \mid A^*=a')\geq 0$ for both $a'$, because $\Pr(Y^{a=1}=1) - \Pr(Y^{a=0}=1 )>0$ by the premise of Proposition \ref{prop: 3} and because Proposition 4 of \citetSM{sarvet2023perspectives} states that if $\Phi_L^*=\Phi_L$ then the sign of the CATEs conditional on $A^*$ are not opposite.  When a CATE is positive or 0 and the conditional probability of \textit{counterfactual} harm is identified, then by Proposition \ref{prop: 2} we know that either $\Pr(Y^{a=1}=1 \mid A^*=a')=1$ or $\Pr(Y^{a=0}=0\mid A^*=a')=1$, which implies that $\Pr(Y^{a=1}=0, Y^{a=0}=1\mid A^*=a')=0$ for both $a' \in \{0,1\}$; that is, in this case the probability of \textit{counterfactual} benefit is 0 for both $a'=1$ and $a'=0$.

If $\Phi_L^*>\Phi_L$, then we know by Proposition 4 of \citetSM{sarvet2023perspectives} that $\Pr(Y^{a=1}=1 \mid A^*=a^\dagger) - \Pr(Y^{a=0}=1 \mid A^*=a^\dagger)>0$ and $\Pr(Y^{a=1}=1 \mid A^*=a^\ddagger) - \Pr(Y^{a=0}=1 \mid A^*=a^\ddagger)<0$ for some $a^\dagger \neq a^\ddagger$, i.e., one CATE is strictly positive and the other is strictly negative. Thus, by the previous arguments $\Pr(Y^{a=1}=0, Y^{a=0}=1 \mid A^*=a^{\dagger})=0$. By similar arguments we can show that $\Pr(Y^{a=1}=1, Y^{a=0}=0 \mid A^*=a^{\ddagger})=0$.  Thus, in this case the probability of \textit{counterfactual} benefit is 0 for  $A^*=a^\dagger$ and the probability of \textit{counterfactual} harm is 0 for $A^*=a^\ddagger$. This concludes the proof.

\bibliographySM{apprefs.bib}
\bibliographystyleSM{unsrtnat_mod.bst}

\end{appendices}


\end{document}